\documentclass[12pt]{article}

\pdfoutput=1 
\usepackage[applemac]{inputenc}
\usepackage[T1]{fontenc}
\usepackage{amsmath}
\usepackage{amsfonts}

\usepackage[english]{babel}

\usepackage{geometry}                
\usepackage[parfill]{parskip}    
\usepackage{graphicx}
\usepackage{amssymb}
\usepackage{epstopdf}
\usepackage[all]{xy}

\usepackage{amsthm} 
 
\theoremstyle{definition} 
 
\theoremstyle{axiom}

\theoremstyle{remark} 
 
\theoremstyle{plain}

\theoremstyle{plain} 

\usepackage{epigraph}
\begin{document}

\title{Are there really many worlds in the "Many-worlds interpretation" of Quantum Mechanics?}
\author{Daniel Parrochia}
\date{University of Lyon (France)}
\maketitle

\textbf{Abstract}
Since the 1970s, the Everett-Wheeler many-worlds interpretation (MWI) of Quantum Mechanics (1955) has been much in the news. One wonders about the "worlds" in question, their "branches", their "splittings", their number. It is most often ignored that this language is not that of Everett, whom Wheeler very quickly stopped supporting. Moreover, for some interpreters, the real meaning of Everett's ideas is not the coexistence of many worlds, but the existence of a single quantum one. In this context, what should we think of attempts to verify Everett's thesis? What about the connexion between MWI and the cosmological multiverse? How to understand the links between the quantum world and the classical one? This article tries to answer some of these questions.\\


\vspace*{\fill} \epigraph{\itshape «He believed in an infinite series of times, in a dizzily growing, ever spreading 
network of diverging, converging and parallel times. This web of time - the strands of 
which approach one another, bifurcate, intersect or ignore each other through the 
centuries - embraces every possibility. We do not exist in most of them. In some 
you exist and not I, while in others I do, and you do not, and in yet others both of us 
exist. In this one, in which chance has favored me, you have come to my gate. In 
another, you, crossing the garden, have found me dead. In yet another, I say these 
very same words, but am an error, a phantom.»}{Jorge Luis Borges\\ (The Garden of Forking Paths (1941))}

\section{Introduction}
 
The interpretation of advanced Quantum Mechanics by Hugh Everett III, in his doctoral dissertation of 1957, is generally presented, since the famous DeWitt article and book of the 1970s (see \cite{Wit1} and \cite{Wit2}) as the «Many-Worlds Interpretation» (MWI).

It is not useless for younger generations of physicists and astrophysicists to recall in what context it was born. As Max Jammer, the famous historian of quantum physics wrote, the first quantum theories of measurement were dualistic, postulating "two fundamentally different modes of behavior of the state functions" and "made the possibility of observation or measurement contingent on the existence of an extraneous macroscopic apparatus or of an ultimate observer" (\cite{Jam}, 508). Now, when the relativist groups at Princeton and Chapel Hill focused their interests on the possible formulations of a quantum theory of general relativity (\cite{Mis}), the idea of a quantization of a closed system like the universe of general relativity, or the idea of a "state of the universe", gradually emerged. But the available dualistic theories of measurement denied to such concepts any operational meanings. So, it was necessary to identify the observer (eventually an automaton) or the measuring apparatus as part of the total system. 

A consequence was to reject the idea of a discontinuous exchange or "collapse" of the wave packet, and this is such a theory which was proposed in 1957 by Hugh Everett III. Then the problem was to explain how we could get, however, a measurement corresponding to what happens in the world we are living.

What Everett proposed is very well summarized by Emily Adlam : «although quantum mechanics predicts that observers should end up in superposition
states, the observer subsystem always has a unique determinate state relative to each distinct state of the instrument subsystem. So if we regard each observer-instrument composite as a single observer, we can understand why measurements always appear
to have a unique outcome.» (see \cite{Adl},4). 

 Everett’s ‘relative state’ formalism has since undergone significant development, and several different
variants have emerged, including the Many-Worlds (\cite{Wit1}, \cite{Wit2}), Many-Minds (\cite{Alb}) and Many-Histories (\cite{Gel} approaches. 
Soon after, the development of «decoherence theory» revealed that, using the standard formalism of quantum mechanics, 
macroscopically distinct branches of the wavefunction were almost entirely free from interference and evolve approximately
classically\footnote{This does not mean that the quantum world is but a joint of classical ones.} (see \cite{Wal1}).

As Adlam also recalls, this has given rise to the Oxford school approach (championed by Deutsch (see \cite{Deu1}, \cite{Deu2}), Saunders (see \cite{Sau}, Wallace (see \cite{Wal2}), Greaves (see \cite{Gre1}) and others), which argues that the quantum worlds may be understood as emergent  features of the structure produced by decoherence, rather than being distinct elements of an ontology (see \cite{Adl}, 4). 

The problem is that the original Everett's thesis never mentioned terms like «worlds», «branches», «minds», «histories» nor any kind of «splitting» of world into worlds. And contrary to what some popularization has claimed (see {\it New Scientist}, 2014, september, the 24th), Everett is not «the man who gave us the multiverse». So, we have to be careful with what is actually said about Everett's theory or its variants (sometimes named Everett-Wheeler or even Graham-Everett-Wheeler MWI). Perhaps we can first recall here what was exactly Everett's thesis.

\section{Everett's thesis, article and comment}

Everett's theory began to be known to the scientific community through the article he published in 1957 (see \cite{Eve1}), but his Ph.D. thesis itself, which he supported at Princeton University, dates back to 1955.

A surprise awaits the reader who opens Everett's thesis for the first time after having heard a lot about it. This one is not called "Many worlds theory" but "The theory of the universal wave function". 

Using at the time the recent von Neumann terminology (see \cite{Neu}), Everett begins by distinguishing two very different processes in quantum mechanics,  the two fundamentally different ways in which the state function can change: 

«Process 1: The discontinuous change brought about by the observation of a quantity with eigenstates $\phi_{1}, \phi_{2},...,$ in which the state $\psi$ will be changed to the state $\phi_{j}$ with probability $|\psi, \phi_{j}|^2$;

Process 2: The continuous, deterministic change of state of the (isolated) system with time according to a wave equation $\frac{\partial \psi}{\partial t}= U \psi$ where $U$ is a linear operator.»(\cite{Eve}, 4).

From this remark, Everett questions the consistency of this scheme where the observer and his object-system form a single (composite) physical system $S$. Indeed, the situation becomes quite paradoxical if one allows for the existence of more than one observer $A$ (\cite {Eve}, 4-6).

Everett then shows very well that we can not escape by limiting the number of observers to one (in fact, they can be several) or by limiting the application of Quantum Mechanics to a domain of experience that would be infra-macroscopic (where would the quantum/classical limit be?\footnote{There is still no decoherence theory at the time. 
The first views about decoherence were introduced by H. Dieter Zeh in 1970.}). Similarly, we can not forbid some observer $B$ to be in possession of the state function of $A + S$, nor to abandon the idea that the state function is a complete description of the system (which would presuppose the introduction of hidden variables).

The only credible alternative is to assume the universal validity of quantum description by abandoning the idea that certain measurement processes would play a privileged role. (In short, Everett proposes to abandon the Process 1, i.e the idea of the so-called «collapse of the wave function» when measuring quantum phenomena). Then:

«Since the universal validity of the state function description is asserted, one can regard the state functions themselves as the fundamental entities, and one can even consider the state function of the whole universe.	In this sense this theory can be called the theory of the "universal wave function",  since all of physics is presumed to follow from this function alone. There remains, however, the question whether or not such a theory can be put into correspondence with our experience.
{\it The present thesis is devoted to showing that this concept of a universal wave mechanics, together with the necessary correlation machinery for its interpretation, forms a logically self consistent description of a universe in which several observers are at work.}» (\cite{Eve}, 9).
 
 Now, in Quantum Mechanics, as we know, a composite system cannot be represented, in general, by a single pair of subsystem states, but can be represented only by a superposition of such pairs of subsystem states. Of course, each element of the superposition may be conceived as containing a definite observer state and a definite relative object-system state.
 
 In this sense, it must be conceded that there is already, in Everett's thesis, what DeWitt will later interpret as a multiplicity of "worlds": as Everett says, «each element of the resulting superposition describes an observer who perceived a definite and generally different result, and to whom it appears that the object-system state has been transformed into the corresponding eigenstate» (\cite{Eve} 10). 
 
 This is clearly expressed in the dissertation. But those "worlds" (the word is not even pronounced), are not worlds in the common sense (with trees, animals, cars, and you and me). They are just observer-system states. Some interpreters like Michel Bitbol (\cite{Bit}) even reduces them to "points of view".
 
 All that constitutes the introduction (Chapter 1) of Everett's thesis. The rest of the dissertation develops his theory in about 80 pages and 4 essential points: 

1) The introduction of some quantitative definitions applying to the attitude of the operators or the degree of correlation of the subsystems of the global composite system; 

2) The mathematical formalization of these notions through the theory of information (Chapter 2: "probability, information and correlation"); 

3) The application of all this to the quantum mechanics of composite systems (the concept of relative state functions, and the meaning of the representation of subsystems by non-interfering mixtures of states characterized by density matrices) (Chapter 3: "Quantum Mechanics"); 

4) The notions of information and correlation are then applied to quantum mechanics (Chapter 4: "Observation"). The final section of this chapter discusses the measurement process, which is simply, for Everett, a correlation-inducing interaction between subsystems {\it of a single isolated system}. A simple example of such a measurement is given and discussed, and some general consequences of the superposition principle are considered. 

The rest of the thesis (about 40 pages) includes two additional chapters.

A fifth chapter (Chapter 5: "Supplementary topics") still discusses some particular situations (relation to macroscopic objects and classical mechanics, the process of amplification, the question of reversibility and irreversibility of measuring processes, the case of approximate measures and, finally, includes a discussion of a spin measurement example). 

Here we do not resist the desire to extract from the point 3 (reversibility and irreversibility) this small passage, particularly edifying for the continuation of our intention :

	«There are, therefore, fundamental restrictions to the knowledge that an observer can obtain about the state of the universe. It is impossible for any observer to discover the total state function of any physical system, since the process of observation itself leaves no independent state for the system or the observer, but only a composite system state in which the object-system states are inextricably bound up with the observer states. As soon as the observation is performed, the composite state is split into a superposition for which each element describes a different object-system state and an observer with (different) knowledge of it. Only the totality of these observer states, with their diverse knowledge, contains complete information about the original object-system state – but there is no possible communication between the observers described by these separate
states. Any single observer can therefore possess knowledge only of the relative state function (relative to his state) of any systems, which is in any case all that is of any importance to him.» (\cite{Eve}, 98-99).

	So, as everyone can see, Everett never said that the reality to which we have access through the quantum wave function could be decomposed into a multiplicity of worlds (which, because of this "decomposition", would be classical). He only said that, "as soon as the observation is performed", the composite state is split {\it into a superposition} of different composite system-observer states.
	
 In a final part (Chapter 6: "Discussion") Everett discusses different interpretations of Quantum Mechanics, including the Copenhagen interpretation. In the latter, Everett blames, in particular, 
 
« its strong reliance upon the classical level from the outset, which precludes any possibility of explaining this level on the basis of an underlying quantum theory. (The deduction of classical phenomena from quantum theory is impossible simply because no meaningful statements can be made without pre-existing classical apparatus to serve as a reference frame.) This interpretation suffers from the dualism of adhering to a "reality" concept (i.e., the possibility of objective description) on the classical level but renouncing the same in the quantum domain.» (\cite{Eve}, 117).

	We would have to remember these two texts when we will study (section 4) the position of the French physicist J.-M. Lévy-Leblond. But, firstable, let us explain why the Everett's interpretation is not, as one believed at the beginning, the «Everett's-Wheeler» theory.

\section{Wheeler's attitude}

	Wheeler's attitude towards Everett and his thesis is a textbook case. At first seduced (it was the time of one spoke, for sure, of the «Everett-Wheeler theory»), he gradually rejected it, and for reasons that do not seem, by far, all rational. Byrne (\cite{Byr}, 307) recalls that, in 1972, Max Jammer, who was writing the last chapter of his {\it Philosophy of Quantum Mechanics} and probably had heard about Everett's thesis, asked Wheeler for his address. By that time, Wheeler had apparently lost track of Everett and could not give him. As Thibault Damour points out (\cite{Dam}, 185), this late request from Jammer proves that he was not aware of this thesis (dating back to 1957) before the echo given it by the DeWitt article of 1970. Wheeler, obviously, had not warned him. 
	
	Later on, Jammer got Everett's address from DeWitt, wrote to Everett, who answered, and finally, was able to devote a few pages of his book to "many worlds theories".

However, Everett seemed, in the meantime, to be at least partly detached from physics. Wheeler, who had first defended him, had been gradually striving to disavow him, probably under the influence of Bohr and the Copenhageners. According to Byrne, as early as the spring of 1957, Bohr had rejected Everett's quantum model as heretic on the grounds that "it violated Bohr's prohibition on the subject of reality as quantum mechanical, as fundamentally non-classical" (\cite{Byr}, 327 ). Wheeler still defended Everett in the {\it Review of Modern Physics} in 1957, but reaffirmed his loyalty to Bohr whom he met at the {\it First Atoms for Peace Award} in Washington DC the same year. Fortunately, Everett was not there to hear the compliments he made to the master. 

As Peter Byrne humorously shows, as long as one only talked about Everett's thesis behind the doors, everything was fine, and Wheeler could still defend it quietly. But "after DeWitt started up his bandwagon, the world at large was paying attention to the Everett-Wheeler theory of multiple universe and Wheeler was torn" (\cite{Byr}, 327). He did not deny that the theory had merit but began to tiptoe away, especially since not only Copenhageners but also physicists like Feynman considered the many-worlds a "ludicrous idea".

Hesitant and cautious, in case the thesis suddenly proved true, Wheeler defended again a certain version of it at the 1972 Trieste Conference, where he claimed that no one could deny the co-existence of alternative stories of the universe in connection with the quantum fluctuations of the geometry of space : as Byrne remarks, it was only the superposition principle writ large. At the same time, however, although in awe of his recent discoveries (the Black Hole concept, a book on gravitation with Misner and Kip Thorne), he seemed increasingly sensitive to relativistic theses of Greek origin on permanent change (ranging from to the point of arguing that the only law in physics is that there is no laws), and now seemed to espouse Bohr's view that the observation changes the observed object. Doing so, Wheeler came dangerously close to Wigner's idealistic positions regarding the fact that consciousness created reality: Everett thought strictly the opposite.

In 1974 in Strasbourg, his religion was made: he now excluded the observer from the wave function, admitted Wigner's argument that the wave mechanics only described correlations between conscious observations and recognized that the wave function of the universe could hardly exist, especially since the adoption of a deterministic universal wave function (as in Everett's dissertation) removed any sense of predictability. Wheeler now described Everett's thesis as extreme and detached himself from it more and more\footnote{Maybe Wheeler had also better arguments, as those exposed in 1979 and reported by Byrne (see \cite{Byr}, 332): for him, Everett's theory included a denial of a really physical quantum character of Nature and a kind of disinterest for a real explanation of the universe where we live.}.

\section{A judicious remark of Lévy-Leblond}

	It is in this context that the French physicist Jean-Marc Lévy-Leblond was led to make a capital remark. Having attended the Strasbourg conference, he wrote a few years later, in August 1977, a letter to Everett which began as follows:
	
	«Dear Dr Everett,
	
	I obtain your address through the kindness of Prof. Wheeler\footnote{Apparently, Wheeler had found it in the meantime...} who suggested that I directly ask your opinion on what I believe to be a crucial question concerning the Everett and no-longer-Wheeler (if I understood correctly!) interpretation of Qu. Mech.»(see \cite{Bar}, 311).
	
	Lévy-Leblond continued as follows :
	
	«The question is one of terminology: to my opinion, there is but a single (quantum) world, with its universal wave function. There are not "many worlds", no "branching", etc., except as an artefact due to insisting once more on a {\it classical} picture of the world. This idea is developed in greater details in pp.184-185 of the enclosed preprint\footnote{The preprint was \cite{Lev}.}.
Your comments on this point, and other ideas in this paper, would be much appreciated, as well as any recent work of yours on this questions.

Sincerely yours, 

Jean-Marc Lévy-Leblond.» (see \cite{Bar}, 312).
	
	As it is said in \cite{Bar}, the Lévy-Leblond's paper from 1976 "concerns the conceptual structure of the quantum measurement problem, the Copenhagen interpretation, and Everett's pure wave mechanics" (\cite {Bar}, 312). Like Everett, Lévy-Leblond blames the Copenhagen interpretation for relying too much on classical intuition, and thus for sacrificing the quantum understanding of the measurement process. But, symmetrically, he blames DeWitt's splitting worlds interpretation of Everett's views which distorts them and make them rely, as in the previous case, on classical world.

In the Copenhagen interpretation, the collapse of the wave function, while projecting the universal state vector, allows to keep only one world in place of the "superposition of states". In Everett's theory, as interpreted by DeWitt, one accepts the simultaneous existence of the many worlds corresponding to all possible outcomes of the measurement. But, in this case too, the "many worlds" idea, according to Lévy-Leblond, is just a "left-over" of classical conceptions.	

	«The coexisting branches here, as the unique surviving one in the Copenhagen point of view, can only be related to worlds described by classical physics. The difference is that, instead of interpretating the quantum "plus" as a classical "or", DeWitt and al. interpret it as a classical "and". To me, the deep meaning of Everett ideas is not the coexistence of many worlds, but on the contrary, the existence of a single quantum one. The main drawback of the "many worlds" terminology is that it leads one to ask the question of "what branch we are on", since it certainly looks as our consciousness definitely belongs to only one world at a time. But this question only makes sense for a classical point of view, one more» (see \cite{Lev} 184-185).
	
	In his response of November 15, 1977, Everett salutes Lévy-Leblond's article as "one of the meaningful papers" that he has read on the subject and gives his full consent to Lévy-Leblond's analysis :
	
	«Dear Prof. Lévy-Leblond,
	
	The reason for the delay in acknowledging receipt of your preprint "Towards a Proper Quantum Theory" is that it is one of the more meaningful papers I have seen on the subject, and therefore deserving of a reply. This is always a mistake for me, as I very rarely complete a thorough review of papers, despite all good intentions. In this case, your observations seem entirely accurate (as far as I read).»
	
	The thing is even clearer when we look at the rough draft of the letter, in which the second paragraph says:
	
	«I have not done further work in this area since the original paper in 1955 (not published in entirely until 1973) as the "Many-Worlds interpretation etc." This, of course, is not my title as I was pleased to have the paper published in any form anyone chose to do it in! I, in effect, had washed my hands of the whole affair in 1955. Far be it for me to look a gift Boswellian writer in the mouth! But your observations are entirely accurate (as far as I have read)» (see \cite{Bar}, 313).
	
	Everett actually says three things in essence: 1) I closed the file in 1955 and all that no longer interests me (I washed my hands); 2) One [DeWitt] wanted to name this theory "many worlds interpretation" and, of course, this is not the title I gave it. But I was not going to be choosy, since one [DeWitt] was republishing my work and, moreover, taking as much interest in myself as Boswell could have done by chronicling the life of Samuel Johnson; 3) That said, you are absolutely right.

	And so it is not questionable, as Barrett (\cite{Bar}, 313) recognizes, that in both version of the letter, Everett {\it agrees} that his formulation of quantum mechanics is best understood in terms of {\it a single quantum world}.

	In other words, this correspondence should not be interpreted as Peter Byrne does, assuming that since the multiple worlds were not actually mentioned in either version of Everett's thesis, Levy-Leblond would probably not understand these were there. In fact, Byrne does not understand the problem (or acts as if he does not understand it) and takes for granted what precisely is in question:
	
	«That the multiverse is a giant superposition embracing branching worlds, some classical, some not, was a key feature of Everett's theory – and this important feature was not clearly explained in either version of his dissertation. It is relevant to why he always viewed the uncountably infinite branches as "equally real".» (see \cite{Byr}, 331).
	
	The fact that Everett's supposed multiverse (with giant superposition embracing branching worlds, etc.) is a key feature of his theory is all the less certain that Everett himself never said that.
	
	Byrne then quotes the end of the letter where Everett explains his disinterest for these problems since 1955 (which is, of course, not quite true), but he carefully omits the point 3 emphasized above, namely that Everett declares its total agreement with the interpretation of Lévy-Leblond!

	In other words, for Everett, exit the famous "many worlds". This term is, in fact, a bad terminology, as Ben-Dov has also well shown in a book directed by Cini and Lévy-Leblond in the 1990s:
	
	«The many-worlds image may perhaps serve as a useful heuristic tool for a first introduction to Everett formulation. But if taken too literally, as is sometimes the case, it leads to consequences which are not necessarily implied by the original mathematical formalism. That is, if one wishes to hold that the measurement process actually “splits” a single world in two (or more), then the splitting “worlds” must be regarded as distinct objects, whose number, at any moment, is well-defined. And as an actual multiplication of things, the “split” itself must be considered as a real physical event, taking place  instantaneously (and at some well-defined moment) at the whole spatial extension of each “world” whenever a quantum measurement is effected. One thus arrives at a quite spectacular world-view, which has probably won for Everett”s theory more publicity than credibility.
	
	But theses features are not necessarily implied by mathematics, and indeed it is quite probable that Everett’s original position was much more loyal to the mathematical formalism than is the “many-worlds” interpretation. In fact, as Everett (1957) is clearly aware, the branches (or for him, simply “elements of the superposition”) appearing in equations [...] are just vectors in Hilbert space. Thus their “number” is representation-dependent, and no objective meaning can be given to their multiplication. In a similar vein, instead of speaking about a world-mutiplying “split” taking place at some well-defined moment, Everett stresses the continuity of the measurement process, which (from the point of view of what he calls “the complete theory”) consists of nothing more than a normal unitary evolution of the composite wave function.
	
	Everett also never hints at every worldwide “split”. For him, the only thing which changes with measurement is the state of the apparatus (or the observer), which becomes correlated with the state of the microobject by interacting with it. It is also significant to note that in both of his published papers Everett (1957, 1973) never uses the term “worlds”, and even the term “split” appears in the article which he originally published (Everett 1957) only in a footnote added in proof, where he cites arguments put forward by “some correspondents” who might have suggested that term in the first place\footnote{As we have seen, however, Everett uses also the word "split" in two occasions in his dissertation (see above and \cite{Eve}, 103-104).}.»(\cite{Ben}, 142).

	In this context, we can probably eliminate the speculations concerning the "number" of the "many worlds" (see, for example, \cite {Hea}), as well as the objections related to them and assuming they are an (uncountable) infinity (see \cite {Esp}; \cite{Pop}, 93; \cite{Gau}, 337-338)\footnote{These questions seem all the less relevant as proponents of the Copenhagen interpretation could also wonder, symmetrically, about the number of times the wave function collapses ...}.
	
	More recently, a very remarkable article by Stephen Boughn noted the difference between the ideas of Everett and his supporters (DeWitt, Graham, etc.). While highlighting the plethora of universes involved in the thesis (not an infinity, however), the author showed both the interest and the problems raised by the position of Everett. Sympathetic with Lévy-Leblond's view, he said to "suspect that most of the conundrums and paradoxes associated with quantum mechanics arise from trying to impose classical views on what are essentially quantum phenomena" (\cite{Bou1}, 14-15).

	Once rid of "many-worlds" fiction and its folklore, knowing that the universal wave function is in fact, for Everett, the only reality, we still have to wonder about this term of "reality", which is not entirely clear. It seems that Everett's realism has been essentially an "operational" realism, so that the wave function is a construct that we invented to deal with our observations of physical phenomena, which brings back its meaning to the use that we do it. But among the different uses, one is to make it settle the question of the measure (and it does not regulate it) and the other to associate it with the whole universe, observer included, which seems to put back (a bit like Spinoza or Hegel) the distinction between theory and what it's about. According to Boughn, and rightly so, we must be again very careful here: with a wave function on the whole Universe, how (and where?) find an experimental / observational corroboration of the theory? In addition, it must be remembered that the "classical cosmology has provided a very successful description of the large-scale structure of the universe without ever worrying that observers are not included in that formalism "(\cite{Bou1}, 28).
	
	For all these reasons, and whatever may be the interest and even the fecundity of Everett's thesis, it should not be considered, for the author, as a good solution, especially since it leaves out what has been at the base of Quantum Mechanics, namely, the quantum of action. Moreover, the question of the collapse of wave function and that of measurement (this latter being also a problem in classical physics) may well be false problems. In this sense, Everett's thesis, rid of its fictional Borgesian aura, leaves us, just like Quantum Mechanics in its usual interpretation, in front of a single world.  		

	But if Everett's thesis is that of a single quantum world, what about {\it experiments} (which remain essentially, until today, {\it thought experiments}) designed to test, not exactly the theory of Everett but the DeWitt distortion of it in the form of the "many-worlds" interpretation? Let's first look at what it's all about.	
	
\section{Testing the Everett's theory?}

In the last part of his first article on Everett's theory ("Final assessment"), DeWitt admitted that «it can never receive operational support in the laboratory» (\cite{Wit1}, 35). No experiment, he said, could reveal the existence of the "other worlds" in a superposition of states like those described in the theory. However, he admitted that a decision between the two main interpretations of Quantum Physics (Bohr and Everett ones) "may ultimately be made on grounds other than direct laboratory experimentation". For example, in the very early moments of the universe, during the cosmological «Big Bang»,  the universal wave function may have possessed an overall coherence as yet unimpaired by condensation into non-interfering branches. Such initial coherence may have testable implications for cosmology.

So, it seems that the Many Worlds Interpretation  (MWI) could be distinguishable from the ideal collapse theory. According to Lev Vaidman (see \cite{Vai2}), the collapse leads to effects that do not exist if the MWI is the correct theory. 

However, as Vaidman has also shown (\cite{Vai2}, \S 5), in order to observe the collapse of the wave function, one would need a super-technology to "undo" a quantum experiment, and in particular to reverse the detection process by macroscopic devices (see \cite{Loc}, 223, \cite{Vai1}, 257) and other proposals by \cite{Deu1}. All of these propositions are in fact «gedanken experiments» that can not be realized with today or forseeable technology. Indeed, in these experiments, it would be necessary to observe an interference of different "worlds". Now, "worlds", in the usual sense, can be judged different if at least one macroscopic object is in states that can be distinguished macroscopically. It is therefore necessary to build an experiment that allows interference with a macroscopic body. Today, there exist interference experiments with larger and larger objects (for example, C$_{70}$ fullerene molecules (see \cite{Bre}). But these objects still do not seem large enough to be considered «macroscopic». Ideally, a truly decisive experience should involve the interference of states that are differentiated by a macroscopic number of degrees of freedom, a task that seems impossible for current technology. It can obviously be argued that the burden falls on the opponents of the MWI, because they claim that there is a new physics beyond the well-tested Schrödinger equation. That said, as shown by Schlosshauer (\cite{Sch}, we have no evidence of this type.

On the other hand, the MWI will prove false if there is a physical process of collapse of the Universe wave function leading to a quantum state in a single world. As one knows, some clever proposals for highlighting such a process have been made (see \cite{Pea}). These proposals – and Weissman's idea of nonlinear decoherence (see \cite{Wei}) could allow us to observe additional effects, such as a tiny non-conservation of energy predicted by several experiments (see for example \cite{Col}). Unfortunately, the effects were not found and some (but not all) of these models are now excluded (see \cite{Adl}).

As Vaidman also notes, much of the experimental evidence of quantum mechanics being statistical in nature, Greaves and Myrvold (see \cite{Gre2}) have done a careful study showing that our experimental data from quantum experiments do not support the MWI probability assumption any more than they take into account Born's rule in other approaches to quantum mechanics. Thus, the statistical analysis of quantum experiments should not really help us to test the MWI. On the other hand, we could evoke a few speculative cosmological arguments that could plead in its favor (see \cite{Pag1}, \cite{Kra}, \cite{Agu} and \cite{Tip}).

 More recently, Aurélien Barrau (see \cite{Barr}), proposed a new experience of this nature. Arguing that the numerous cosmological multiverses are testable, the author deduces that this must also be the case for the quantum many-worlds. Concerning the test for MWI, some authors already made certain proposals. Until now, three have been considered :
 
 1. The first test is connected to linearity. Because of linearity, it is possible to detect in some world the presence of other nearby worlds, through the existence of interference effects. However, this may be nearly impossible to  experiment.
 
 2. A second test implies gravity. The Many-worlds theory requires that gravity be quantized. Now, if gravity was to remain non-quantized, all the universes that the Everett interpretation predicts should become detectable (by their gravitational presence) – and so this detection would falsify the theory. Of course, this test is not crucial because gravity could be quantized and the Everett theory still be wrong.

3. The third one, related with the first, is based on reversible quantum computers. But it is today beyond the reach of artificial Intelligence.

Barrau proposes to model the many-worlds theory test on the cosmological multiverse test, as, for example, Page (see \cite{Pag1}). The idea is that, in the case of a single universe, if the probability of existence of a $X$ universe is greater than that of a $Y$ universe, then we must certainly be in the $X$ universe. However, in DeWitt's design, all possible universes really exist. In this case, it is necessary to involve the number of observers to decide between them. If the numbers of observers - respectively $N_ {X}$ and $ N_{Y}$ are different in the universes $X$ and $Y$, the probabilities of existence of these worlds are weighted by the number of observers in each of them. For example, if the ratio $N_ {Y} / N_ {X}$ is greater than the ratio $P (X) / P (Y)$, then we will be in the universe $Y$ instead. As Barrau writes, "the point is that the situation is basically the same as in any multiverse situation. If there in only one World, we should compute probability for this World, if there are many worlds the observer-weighted probability is the correct distribution. In principle, observations can select which one is true" (\cite {Barr}, 131).

The problem, obviously, is to fix the number of observers of the universes. Reasoning on the Hartle-Hawking no boundary proposal, Page showed that the no boundary proposal led to a set of FLRW universes with a varied number of inflationary leaflets and different properties. In this context, the amplitude of probability of these worlds seemed proportional to the size and volume of the universes and became also a coarse measure proportional to the number of observers.

Then "one can compare the probability for our Universe being what it is in a single universe history on the one hand and, on the other hand, with the observer-weighted probability that should be chosen in the Everett many-worlds view where all universes do actually exist. The result strongly favors the Everett view" (see \cite{Barr}, 132).

Of course, as Barrau recognizes, this result should not be taken too seriously : Hartle-Hawking proposal may be wrong and the result of Page relies on many controversial assumptions. But this shows that testing the many-worlds theory is possible and the approach can be generalized.

It remains that the multiverse, coming from cosmological speculations, and the so-called quantum «worlds» (in fact, superposed state vectors coming from quantum physics MWI), initially refer to completely different fields of knowledge, very difficult to connect, even if what brought them together seems well-known.

Everybody remembers that, after the introduction of inflation in cosmology by Starobinsky, Guth and Linde in the 1970s and early 1980s, the main proposed ingredient of the putative mechanism to explain it was the so-far unidentified field, called the {\it inflaton}. But as there were many models with different potentials for the {\it inflaton}, each was supposed to give rise to a multiverse, i.e to a vast set of non-interacting buble universes - of which one is our own universe.These bubble universes varied, not just in details, but also in the values of fundamental physical parameters such as the cosmological constant or the fine structure constant. All these formed the cosmological multiverse. 

Here there has been a confluence with developments in the last twenty years within string theory. Back in the 1980s, string theorists hoped the constraints on constructing a consistent string theory would be so strong that there would lead to a unique consistent big theory or, at least, only a few such. But, as \cite{But} said very well,  this hope, as we know, has been dashed. 

	"In the last twenty years, it has turned out that string theory admits a vast number of local ground states (metastable vacua): states that are in a local minimum of the potential. A truly vast number: an estimate often cited is $10^{500}$, while Taylor and Wang’s (2015) estimate is $10^{272,000}$ – daunting, indeed depressing, numbers. The ‘towers’ of excited states built up from each such ground state would then count as different string theories. This is the string theory landscape. Besides, the overall scheme of string theory suggests that the different theories – the towers of excited states – differ in values of fundamental physical parameters, such as the cosmological constant and the fine structure constant.  So here is the confluence with the cosmological multiverse. For in both cases, we are confronted with a vast population of what one might (to choose a neutral word) call ‘domains’: domains that vary in fundamental parameters. Furthermore, this confluence is strengthened by a widespread, albeit usually implicit, adoption by string theorists of the Everett interpretation of quantum theory. For the idea of the cosmological multiverse is that the countless domains are all equally real. (...) But according to the Everett interpretation, one can say the same about the various towers of excited states built up from the various vacua in the string theory landscape. That is: suppose that the quantum state of the entire cosmos is some sort of sum, or integral, of states in the various different towers (or a sum of tensor products of such); or is a mixture, i.e. density matrix, with these as components. Then there is, in Everettian parlance, an amplitude for various branches associated with various different vacua: or more precisely, associated with excitations above various different vacua. So there is an amplitude for various different values of fundamental parameters. And according to the Everettian, the different branches are equally real: just as, we saw, the countless domains of the cosmological multiverse are meant to be" (see \cite{But}, 14-15).

 Such hypotheses generate a lot of problems. But of course some physicists pretend to  ignore them and, without looking further, identify the two situations altogether. A fascinating article by Nomura tries a unification (\cite{Nom},36) and concludes "that the eternally inflating multiverse and many worlds in quantum mechanics are the same" (see also, on the same theme, the article by Bousso and Susskind \cite {Bou2}). The less one can say is that this remains to be confirmed, especially as Nomura uncritically takes up DeWitt's image of the different branches (see the picture of the tree in \cite{Nom}, 37). In general, there is still a lot of instability in these speculations, some wanting, for example, to build an interpretation of the MWI that allows to deduce the Born's rule (see \cite{Aer}), while others do not hesitate, on the contrary, to announce its definitive death (see \cite{Pag2}).

This is in addition to the problems posed by DeWitt's interpretation, and recently well summarized by Tappenden in (\cite{Tap}): The MWI, from the beginning, already asks a lot of questions.The moment in which the branching of the worlds intervenes as the form that this one takes is not clear: one can imagine several models of branching and roads overlapping partially one on the other (see \cite{Tap}, 123). The associated probabilities are also problematic. Moreover, Maudlin (\cite{Mau}) was able to point out that, as soon as  there is a wavefunction, an extremely high-dimensional object evolves in some specific way. So the question is: how does it account for the low-dimensional world of localized objects? These questions do not seem to have received definitive answers yet and the opinions of the physicists, on all these points, diverge.

In short, Everett's thesis, as reviewed and corrected by DeWitt, lends to science-fiction. But we must not forget that there is science under fiction and that much of many-worlds theory is a "way of speaking". 

	Going further risks exposing physicists to the harsh criticisms of Sabine Hossenfelder who argues that the physical community has wandered and that, for a long time, physics has taken a wrong turn with this idea of "multiverse":
		
	"The	multiverse has	gained in popularity 	while naturalness, has come under stress, and	physicists	now	pitch	one as the other’s alternative. If we	can’t	find	a natural explanation for	a number,	 so the argument goes, then there isn’t any. Just choosing a parameter is too ugly.	Therefore,	 if the parameter is not natural,	then	it can take	on any	value, and for every	possible value	there’s a universe. This leads to the	bizarre conclusion that if we don’t see supersymmetric particles at the LHC, then we live	in a multiverse.	
	I can’t believe  what	this once-venerable	profession has	become. Theoretical physicists	 used to explain what was	observed.	Now	they	try to	explain why they can’t explain	what	was	not	observed.	And	they’re not even good at that. In a multiverse, you can’t explain the	values of parameters; at best you can estimate	their	likelihood.	 But there	are	many ways not	to explain	something." (\cite{Hos}, 107-108).

\section{Logic of inconsistency and Lévy-Leblond's criticism}

As has been shown in section 4, it is clear, for us, that the theory of multiple worlds is essentially a communicative effect, a kind of media formulation of Everett's theory due to the fanfare of DeWitt's writings. In this context, we would do well not to take it literally, except, as Lévy-Leblond showed, to make sneaky shifts which, finally, tend to re-establish a classical language to interpret or describe something that is clearly non classical.

But the questions that can be asked of Lévy-Leblond are the following: how could the human mind claim to "understand" quantum mechanics without bringing it back to elements directly interpretable in our common language? By remaining as close as possible to the mathematical formalism of Hilbert spaces and state vectors, without any sketch of possible translation, do we not close, on the contrary, any issue to really "understand"? Or must we think this formalism is not absolutely separated from classical one?

One way out could be to take a glance at some non classical logics.

The situation described by Lévy-Leblond as a characteristic of the duality between the Copenhagen interpretation of Quantum Mechanics and the DeWitt's interpretation of Everett's theory recalls the Brandom-Rescher "logic of inconsistency", introduced in 1980.

The Brandom-Rescher logic of inconsistency is a study in Non-Standard possible-worlds semantics and ontology. The authors introduced a modal semantics including, besides ordinary possible worlds (taken as maximally consistent collections of states of affairs), also non-standard worlds that are locally inconsistent or incomplete.

Inconsistent worlds are such that, for some proposition $A$, both $A$ and $\neg A$ hold at them. Incomplete worlds are such that, for some proposition $A$, neither $A$ nor $\neg A$ hold at them.

An important point is that these nonstandard worlds are combinatorially generated from standards worlds via two recursive operations having standard worlds at there base. These ones are  called superposition ($\dot{\cup}$) and schematization ($\dot{\cap}$).

Given two worlds $w_{1}$ and $w_{2}$, a schematic world $w_{1} \dot{\cap} w_{2}$ is one at which all and only the states of affairs obtain, which obtain both at $w_{1}$ and at $w_{2}$. Dually, a “superposed” or inconsistent world $w_{1} \dot{\cup} w_{2}$ is one at which  all and only the states of affairs obtain, which obtain at $w_{1}$ or at $w_{2}$.

 According to Norton, this could perfectly apply in physics, to solve physical inconsistencies and to explain the relationships between classical and quantum physics : "This device enables logical inconsistency to be tolerated without anarchy and provides the kind of rule needed to govern exchanges of propositions between the classic and quantum domains"(\cite{Nor}, 330-331). 
 
 But it is precisely what Lévy-Leblond refused.

The situation is irresistibly reminiscent of the French physicist. The DeWitt many-world theory that interprets the operation "+" adding 2 state vectors as the logical "and" connector of logic is like settling into an inconsistent world. While the Copenhagen interpretation, that induces a simplification of the world by admitting the collapse of the wave function installs the observer in a kind of incomplete world. The interpretation of the connectors is reversed, but it is the purpose of the formalism to oppose inconsistency and incompleteness.

So, we might think that the Rescher-Brandom logic of inconsistency does not really formalize the situation encountered in quantum mechanics. It just shows more clearly why some well-known interpretation like the Copenhagen interpretation or the DeWitt many-worlds theory misses the specificity of the situation : inconsistent or incomplete worlds are just combination of classical ones. The problem is that the examination of actual quantum logic does not lead to a quite different result.

\section{Physics and quantum logic} 
	
	What about the canonic forms of quantum logic?
	
	There are certainly many ways to consider quantum logic\footnote{Quantum logic can be formulated either as a modified version of propositional logic or as a noncommutative and non-associative many-valued logic and there has been a number of different approaches (\cite{Jam}; \cite{Eng}).} and we can not examine even a very small part of these many models. But it is well known that one of the most prominent is Birkhoff and Von Neumann formalization (see \cite{Bir}; and, more recently \cite{Jam}, 341-416), which is entirely derived from the Hilbert spaces of quantum mechanics. So let us briefly expose it.
	
	In this theory, all logical primitives, such as propositions, logical implication relation, operators "and", "or", "no", must refer to entities in the Hilbert space. The semantic values "true" (1) or "false" (0) are attributed to the presence or absence of a particular property of these entities (for example, the fact that a quantity $R$ takes the value $\lambda$ or that the value of $R$ is positive, etc.). We can then show that the properties correspond to the subspaces of the Hilbert space.

	Indeed, in quantum mechanics, as we know, an observable is represented by a self-adjoint operator $A$, that is to say such that $A = (Ax, y) = A^{\dagger} = (x, Ay)$ which can have a decomposition. 

	Now let $\mathbf{E}$ be a property whose corresponding operator is $E$. If, in the polynomial $F (\lambda) = \lambda - \lambda^2$, we replace $\lambda$ with $E$, then we obtain $E = E^2$, in other words, we define the operator $E$ of $\mathbf{E}$ as a {\it projection}.

	The eigenvalues $\lambda_{1}, \lambda_{2}, ..., \lambda_{r}$ therefore correspond to orthogonal projection operators $E_{1}, E_{2}, ..., E_{r} \ (\neq 0$) with $r \leq n$ such that:

\[
A = \sum_{i=1}^r \lambda_{i}E_{i},
\]
with:
\[
\sum_{i=1}^r E_{i} = \mathbf{H}.
\]

The $\lambda_{i}$ giving the measures, the projection operators E are in correspondence with elementary propositions of the type "the system is in the state $i$" or "the measurement outcome is $\lambda_{i}$, i.e. propositions that can be true (value 1) or false (value 0) (see \cite{Svo}, 7-9).

Measurements taken together thus lead to combinations of orthogonal projection operators and, consequently, to combinations of propositions. The set gives rise to particular lattices, sometimes quite different from the lattices of classical mechanics (see \cite{Hug1}, \cite{Hug2}).

\subsection{The case of spin one-half}

Take, for example, the case of spin one-half (we follow here \cite{Svo}, 26-28). Consider measurements of the spin-component along one particular direction, say along the $x$-axis (which can be operationalized with a Stern-Gerlach type experiment). There are two possible spin (angular momentum) components of the particle, namely $-\frac{1}{2}, +\frac{1}{2}$ (in units of $\hbar$). We will say that the particle is in state - if it has spin $-\frac{1}{2}$ and in state + if it has spin $+\frac{1}{2}$. This corresponds to the following elementary propositions:

Let $\mathbf{H}$ be the 2-dimensional, real Hilbert space $\mathbb{R}^2$ with the usual scalar product $(v, w) := \sum_{i=1}^2 v_{i}w_{i}$.

\begin{enumerate}
\item Any proposition is identified with a closed subspace of $\mathbb{R}^2$.
\item The zero-vector corresponds to a false statement.
\item  The entire Hilbert space $\mathbb{R}^2$ corresponds to a tautology.
\item Any line spanned  by a nonzero vector corresponds to the statement that the physical system has the property associated with the closed linear subspace spanned by the vector.
\end{enumerate}

$p_{-}$ : The particle is in the state -.

$p_{+}$ : The particle is in the state +.

1 : The particle is in the state - or + (entire Hilbert space).

0 : The particle is neither in state -, nor in state + (zero-dimensional subspace).

As the complement of a line is a line orthogonal to that line, and as the complement of the zero-dimensional subspace is the entire Hilbert space (and vice-versa), we have the following structure (boolean lattice) with :
\[
 p_{-} = p'_{+}, p_{+} = p'_{-}, 1 = p_{-} \vee p_{+} = 0' = (p_{-} \wedge p_{+})'.
 \]
 This structure, concerning comeasurable observables is classic and here the distributional laws are satisfied. 
 
 Now consider the following more complex situation where we get two noncomeasurable systems. Let $L(x)$ such that :
 
 \[
 L(x) =  \{0, p_{-}, p_{+}, 1\}
 \]
 
 a propositional system corresponding, as above, to the outcomes of the measurement of the spin states along the $x$-axis, and let also:

 \[
 L(\bar{x}) =  \{\bar{0}, \bar{p}_{-}, \bar{p}_{+}, \bar{1}\}
 \]
 
be another propositional system corresponding to the outcomes of the measurement of the spin states along a different spatial direction, say $\bar{x} \neq x$ mod.$\pi$. $L(x)$ and $L(\bar{x})$ can be jointly represented by pasting them together. Tautologies and absurdities being identified ($0 = \bar{0}, 1 = \bar{1}$), all the other propositions remain distinct and we then obtain the propositional structure $L(x) \oplus L(\bar{x})$, whose Hasse diagram is the "Chinese lantern" of Fig. 1.

\begin{figure}[htbp] 
	   \centering
	      \vspace{-3\baselineskip}
	   \includegraphics[width=4in]{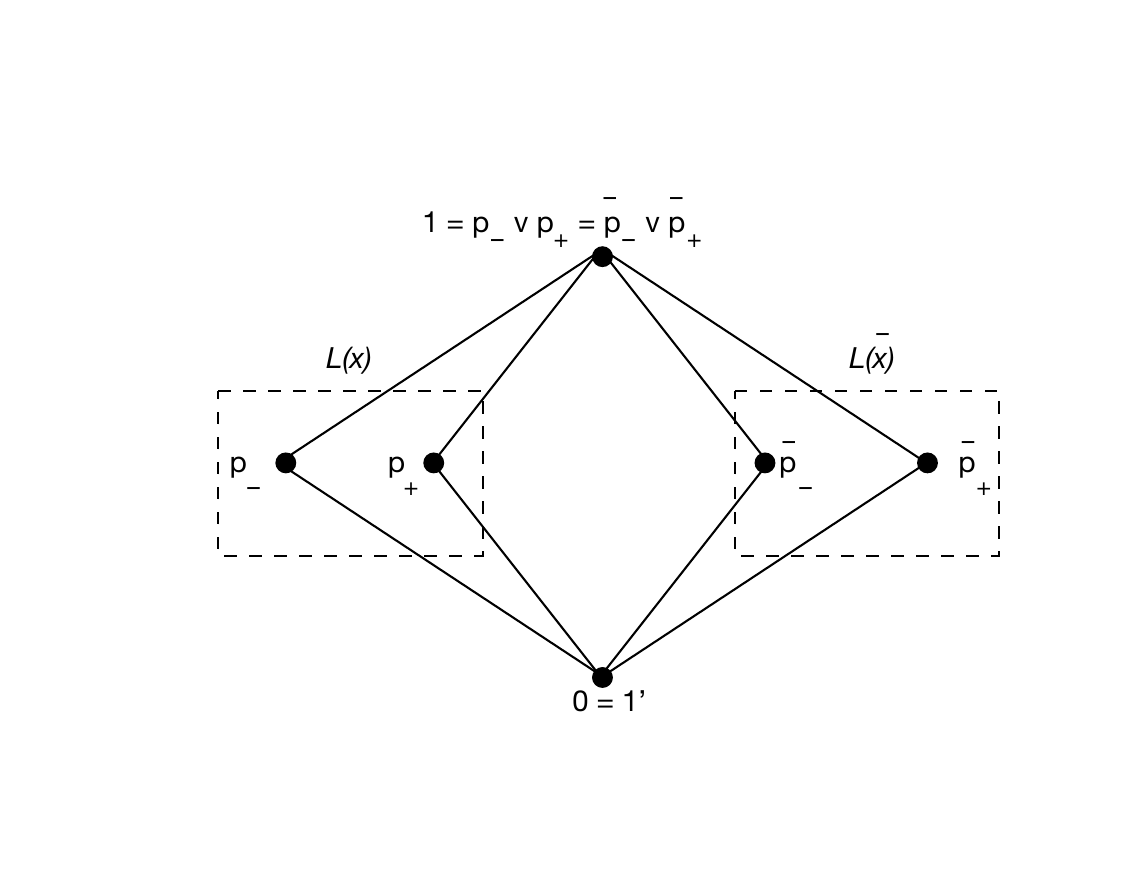}
	      \vspace{-3\baselineskip}
	   \caption{Hasse diagram for 2 spin one-half non comeasurable systems}
	   \label{fig: Hilb1}
	\end{figure}

	It is well known that such a lattice is not distributive, which can be easily proved (see \cite{Svo}, 27-28). Moreover, since every element has a complement, it is an orthocomplemented lattice. Finally, one proves this lattice is also modular, because the modular law $(a \vee b) \wedge c = a \vee (b \wedge c)$  is always satisfied. Then we can call this lattice $MO_{2}$ (M for "modular", O for "orthocomplemented", the subscript "2"  standing for the pasting of the two Boolean subalgebras $2^2$. So we have :
	\[
MO_{2} = L(x) \oplus L(\bar{x}),
\]
	a structure that can be generalized in:
	\[
MO_{n} = \oplus_{i = 1}^n L(x^j),
\]
if we consider a finite number $n$ of different directions of spin state measurements. As written above, the resulting structure is the horizontal sum $MO_{n}$ of $n$ classical Boolean algebras $L(x^j)$, where $x^j$ indicates the direction of spin measurements.

\subsection{Experiments in a 3-dimensional Hilbert space} 
Let us see now another example. Consider the situation described by Foulis and Randall (see \cite{Fou}) which is a realization of a 3-dimensional case. Be $D$ a device such that, from time to time, it emits a particle and projects it  along a linear scale.Two types of experiments ($A$ and $B$) are performed. In Experiment $A$, we look to see if there is a particle present. If there is no particle present, we record the outcome of $A$ as the symbol $n$. If there is a particle present, we measure its position coordinate $x$. It $x \ge 1$, we record the outcome of $A$ as te symbol $r$; otherwise, we record the symbol $\ell$. Similarly, for experiment $B$ : if there is no particle, we record the outcome of $B$ as the symbol $n$. If a particle is present, we measure the $x$-component $p_{x}$ of its momentum.If $p_{x} \ge 1$, we write $b$ as for the outcome, otherwise we write $f$ (\cite{Svo}, 34). 

The resulting lattice is named $L_{12}$ (or sometimes $G_{12}$) has 12 elements and its logical structure, in 3-dimensional Hilbert space, corresponds, as Svozil says, «to two tripods glued together at one leg», in fact two triples of orthogonal vectors whose some points has been identified (see Fig. 2).

\begin{figure}[h] 
	   \centering
	      \vspace{-3\baselineskip}
	   \includegraphics[width=4in]{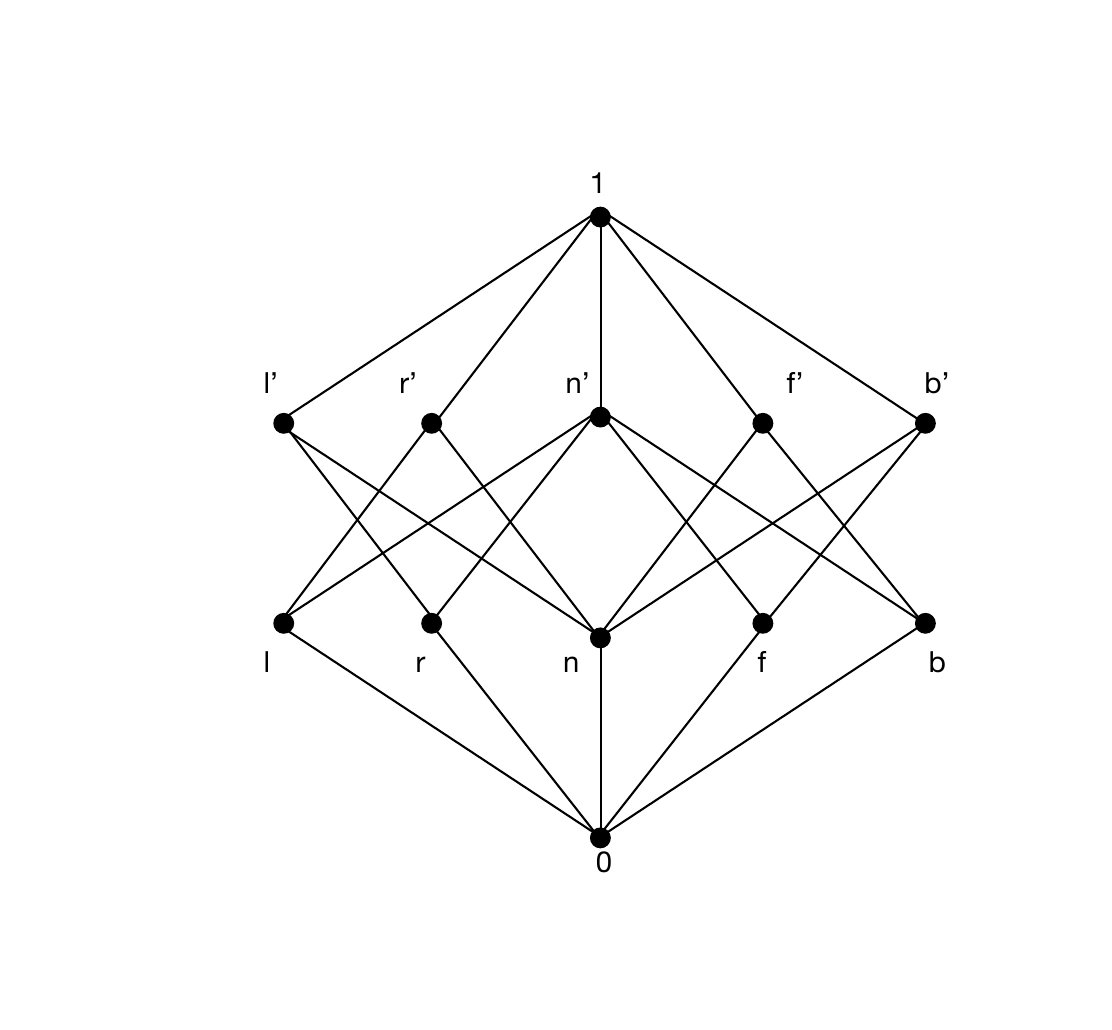}
	      \vspace{-3\baselineskip}
	   \caption{Hasse diagram of the lattice $L_{12}$ in 3-dimensional Hilbert space}
	   \label{fig: Hilb1}
	\end{figure}

This lattice, which is not a sum of subalgebras, as previously, is in fact the product of a Boolean algebra of dimension 2 and a modular orthocomplemented lattice $MO_{2}$ (see Fig. 3).

\begin{figure}[h] 
	   \centering
	      \vspace{-2\baselineskip}
	   \includegraphics[width=6in]{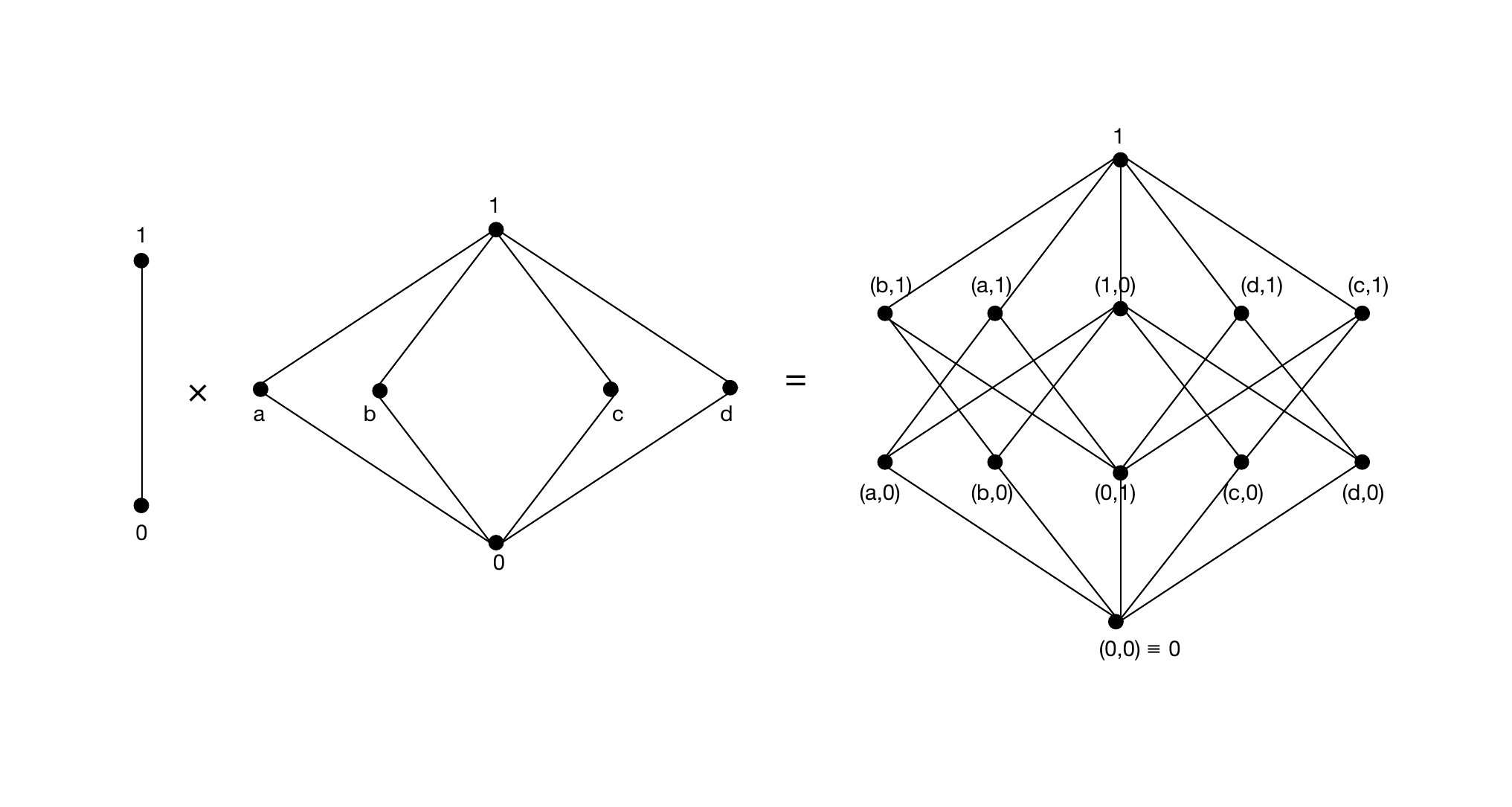}
	      \vspace{-4\baselineskip}
	   \caption{Cartesian product $2^1 \times MO_{2} = L_{12}$}
	   \label{fig: Hilb1}
	\end{figure}
	
This logical structure, discussed by Foulis and Randall can be easily generalized (see \cite{Svo}, 36-39). With the same type of device as above (the case of spin-one half particle projected along a linear scale), we now perform $n$ types of experiments labelled by $i$. Every $i$ is associated with a direction of angular momentum measurement $x^i$. In the $i$th type of experiment, we look to see if there is a particle present. If not, we record the outcome of $i$ as the symbol $n$. If there is a particle, we measure its spin state. If it is in state –, we record the outcome of $i$ as $p^i_{-}$, otherwise, we record $p^i_{-}$. 

The previous results can be generalized to $n$-dimensional Hilbert spaces. In dimension $n$, the resulting lattice would be the product of a Boolean algebra of dimension $n-2$ and a modular lattice of the type "Chinese lantern" denoted $MO_{m}$, such that:

\[
B \times MO_{n} = 2^{n-2} \times MO_{m}, \qquad \quad 1 < m \in \mathbb{N}, \quad n\ge 3.
\]
In general, we have:
\[
L_{2(2m+2)} = L_{4m+4} = 2^1 \times MO_{m}.
\]
Apparently, nondistributivity, modularity and orthocomplementarity characterize quantum lattices, and seem to be specific properties of quantum logic. However, when we pay attention to how these structures are produced, we discover that they are generated by summing classical Boolean structures or gluing mixed Boolean structures and orthocomplemented modular ones. In the first case, quantum logic is just an extension of the classical situation. In the last case, we could have identified something new if orthocomplemented modular structures could not decompose into boolean ones. But the fact is it decomposes, and the cartesian product that generate the resulting lattice is a quite classical structure.

So it seems that quantum logic also fails to grasp the originality of the new physics since it brings it back to a simple combinatorial of classical structures (not "worlds" as in Rescher-Brandom logic but Boolean lattices). More recent works in quantum logic, and in particular those that refer to the so-called "dynamic turn" of quantum logic (see \cite{Bal}) continue to go in this direction.

In an article already quoted, Michel Bitbol (\cite{Bit}), who has himself perceived the contradiction between the orthocomplemented lattices of quantum mechanics and the perspectivism of Boolean representations of experiments, is doing feats to save both quantum originality and Everett's theory. This one, even though it seems to take into account the realism of the wave function, can only be resolved, at less in the DeWitt presentation, in the perspectivism of classic multiple worlds. Solving the contradiction forces  finally Bitbol to wonder what an observer really is (for him, essentially an "abstract mind", so that his interpretation of Everett's theory relies both on an extended version of the perspectival definition of "real objects", and also on a limiting concept of the observer, namely a pure knowing subject).

But the problem is not only to find a philosophical solution, but before all, a physical solution. 

The consequence of all this is that the arguments of Lévy-Leblond or Boughn in favour of a single quantum world must be precised, as it appears that quantum reality, even globally and specifically understood, can always split into a plurality of classical structures.  Of course, it probably does not mean that these are the multiple branches of a real cosmological tree and that the quantum world is a multiverse similar to the borgesian «garden of forking paths» with many "I" (or "you") and, among them, some phantoms.  But this reality, both global and split, remains in fact rather mysterious : is it real? Is it only in our mind? Does it have a reality only temporary or transitory? Is it, for lack of a better, a provisional representation? Nobody knows, and the theory of universal wave function, contrary to what one might think if we follow the popular presentation of DeWitt, does not clarify much.

\section{Conclusion}

	Commenting on Putnam's realism, Yannis Delmas-Rigoutsos (\cite{Rig}, 19-20) shows that, for getting a good logic reflecting the situation in Quantum Mechanics, we must weaken certain realistic propositions and change them into others classically equivalent but true from a quantic point of view. 
	
	It is relatively easy in the case of the Heisenberg uncertainty principle where, as in the case of spin, the non comeasurable variables (here pulse and position) can not satisfy distributive laws, we are dealing with some proposition of the type: "S simultaneously has a definite position and definite impulse" or another one which is classically equivalent to it "S has definite position and impulse". But we know that these two assertions are formalized very differently. The first :
\[
(x_{1} \wedge p_{1}) \vee ... v\vee (x_{i} \wedge p_{i}) \vee ... \vee (x_{n} \wedge p_{m}),
\]
is wrong in Quantum Logic while the second:
\[
(x_{1} \vee ... \vee x_{n}) \wedge(p_{1} \vee ... \vee p_{m})
\]
 is true in this very logic.	
	
	 The case of the double split experiment may be treated in a similar way. Let $A_{1}$ be the statement : "the photon passes through slit 1", and let $A_{2}$ be the statement "the photon passes through slit 2". Let the probability that the photon hits a tiny region $R$ on the photographic plate on the assumption $A_{1}$ be $P(A_{1}, R)$, and let the probability of hitting $R$ on the assumption $A_{2}$ be $P(A_{2}, R)$. These probabilities are the same in classical or quantum mechanics when only one slit is open. Now, if both slits are open, the probability $\mathcal{P}$ that the particle hits $R$ as predicted by classical mechanics is :
	
	\[
	\mathcal{P} = \frac{1}{2} P(A_{1},R)+\frac{1}{2} P(A_{2},R).
	\]

	But it is not what is observed, and indeed is correctly predicted by quantum mechanics. As Putnam shows, the crucial point in the derivation of the classical prediction is that, in accordance with classical logic, one has expanded $(A_{1} \vee A_{2}) \wedge R$ into $(A_{1}  \wedge R) \vee (A_{2}  \wedge R)$, which is fallacious in quantum logic. Putnam does not solve the problem :
	
	«Someone who believes classical logic must conclude from the failure of the classical law that one photon can somehow go through two slits (which would invalided the above deduction, which relied at many points on the incompatibility of $A_{1}$ and $A_{2}$), or believe that the electron somehow "prefers" one slit to the other (but only when no detector is placed in the slit to detect the mysterious preference), or believe that in some strange way the electron going through slit 1 "knows" that slit 2 is open and behaves differently than it would if slit 2 were closed; while someone who believe quantum logic would see no reason to predict $P(A_{1} \vee A_{2}, R) = \frac{1}{2} P(A_{1},R)+\frac{1}{2} P(A_{2},R)$ in the first place.»(\cite{Put}, 181).
	
	So we have a real problem to translate into natural language and make understandable to a human mind what the formulas of quantum mechanics say. This has philosophical consequences.
	
	In the 18th century, the English philosopher G. Berkeley thought that "to be is to be perceived". Today Quantum Mechanics no longer allows us to leave such an important place for phenomena, since at no time can phenomenal knowledge be complete: the God of Berkeley, like all omniscient beings in the strong sense, can not exist. Similarly, Kant's "thing in itself" can no longer exist. Quantum physics shows us that the so-called "in-itself" can not be governed by classical logic because it conveys at all times all future potentialities, which include incompatible elements (see \cite{Rig}, 21). On the other hand, the phenomenon, at least if one follows the common sense, is based on human experience, and thus includes, at certain moments, elements undecided. So this  kind of experience is necessarily incomplete and its description, which, therefore, does not include incompatibilities, can be based on classical logic. But with modern science, our description of reality comes to be refined. We can no longer base this description on a series of values of truth assigned to a series of given common sense propositions. Quantum Mechanics founds us to recognize that the scientific world is only, as the French philosopher G. Bachelard (\cite{Bac}, 11) said, our {\it verification}, i.e. the sanction of reality. In this context, quantum reality is our reality, quantum logic is our logic. But human mind continues to decompose all that in order to understand, and it is very difficult to find a true language between mathematics and fairy tales.

\end{document}